\documentclass[11pt]{article}
\usepackage[margin=1in]{geometry}
\usepackage{graphicx}
\usepackage{tabularx}
\usepackage{hyperref}
\usepackage{caption}
\usepackage{subcaption}
\usepackage{amsmath}  
\usepackage{amssymb}  
\usepackage{booktabs}
\usepackage{tabularx}
\usepackage{verbatim}
\usepackage{pgfplots}
\usepgfplotslibrary{fillbetween}  
\usepackage[table]{xcolor}
\usepackage{multirow}
\usepackage{float}
\usepackage{color, soul}
\sethlcolor{yellow}
\usepackage{algorithmic}
\usepackage{xcolor}
 \usepackage{cleveref} 
\usepackage{algorithm}
\newcolumntype{M}[1]{>{\centering\arraybackslash}m{#1}}
\usepackage{tabularx}
\usepackage{enumitem}
\usetikzlibrary{intersections}
\usepackage{booktabs}

\newcommand{\vx}[0]{{\bf x}}

\newcommand{\vy}[0]{{\bf y}}

\newcommand{\bmx}[0]{\begin{bmatrix}}
\newcommand{\emx}[0]{\end{bmatrix}}

\newcommand{\samplesize}{N}
\newcommand{\sampleidx}{i}

\newcommand{\actionitem}{y}

\newcommand\labelvec{\vy}
\newcommand\featurevec{\vx}

\newcommand\predictedlabelvec{\hat{\labelvec}}

\newcommand\realcoorspace[1]{\mathbb{R}^{#1}}

\newcommand{\sigmoid}{\sigma}

\newcommand{\loss}[3]{\mathcal{L}_{\text{\tiny{#1}}}\left({#2},{#3}\right)}

\newcommand{\timeidx}{t}

\newcommand{\activityidx}{j}                          
\newcommand{\nrsubcarriers}{N_\text{sc}}
\newcommand{\subcarrieridx}{k}
\newcommand{\numActivities}{N_\text{act}}
\newcommand{\numqueries}{N_\text{q}}

\newcommand{\numsubcarriers}{N_\text{sc}}
\newcommand{\initLength}{T}
\newcommand{\initChannel}{C}

\newcommand{\numtxant}{N_\text{t}}                       
\newcommand{\numrxant}{N_\text{r}}                       

\newcommand{\TrueNumPeople}{N_\text{p}}
\newcommand{{\replength}}{T_\text{z}}

\newcommand{\setofPermutations}{\mathcal{P}}

\newcommand{\numEncoderLayers}{L_\text{E}}              
\newcommand{\numDecoderLayers}{L_\text{D}}              
\newcommand{\numAttentionHeads}{N_\text{h}}
\newcommand{\embedDim}{D_\text{z}}                        
\newcommand{\keyDim}{d_\text{k}}                        
\newcommand{\encoderInput}{\hat{\mathbf{B}}}          
\newcommand{\encstate}[1]{\hat{\mathbf{B}}^{(#1)}}   
\newcommand{\encattn}[1]{\mathbf{O}^{(#1)}}          
\newcommand{\encoutput}{\mathbf{H}}                   
\newcommand{\numRVQLayers}{V}                    
\newcommand{\codebookSize}{\kappa}                    
\newcommand{\commitmentCost}{\beta}              

\newcommand{\learningRate}{\alpha_{\text{lr}}}   
\newcommand{\numEpochs}{N_\text{ep}}                       
\newcommand{\batchSize}{N_\text{bc}}                       
\newcommand{\weightDecay}{\lambda}               

\newcommand{\kernelSizeDSC}[1]{{k^{(#1)}_{\text{d}}}}      
\newcommand{\kernelSizeAC}{{k_{\text{a}}}}        
\newcommand{\strideParam}[1]{{s_\text{d}^{(#1)}}}                     
\newcommand{\dilationRate}[1]{{d_\text{a}^{(#1)}}}                    
\newcommand{\numOutputChannelsDSC}[1]{{C^{(#1)}_{\text{d,out}}}}  
\newcommand{\numOutputChannelsAC}[1]{{C^{(#1)}_{\text{a,out}}}}  

\newcommand{\smoothLossBeta}{\beta_{\text{smooth}}} 

\newcommand{\queryemb}{{\mathbf{Q}_{\mathrm{act}}}} 
\newcommand{\decstate}[1]{{\mathbf{U}^{(#1)}}} 
\newcommand{\decstatecross}[1]{{\mathbf{U}_{\mathrm{cross}}^{(#1)}}} 
\newcommand{\decstateself}[1]{{\mathbf{U}_{\mathrm{self}}^{(#1)}}} 
\newcommand{\auxLossCoef}{\alpha_{\text{aux}}}
\setlist[itemize]{leftmargin=*}

\setlist{leftmargin=*}

\newenvironment{IEEEkeywords}{\vspace{0.5em}\par\noindent\textbf{\textit{Index Terms}---}}{\par\vspace{0.5em}}

\title{AMAR: Lightweight Attention-Based Multi-User Activity
Recognition from Wi-Fi CSI}

\author{
  Amirhossein Mohammadi \qquad Hina Tabassum \\[0.4em]
  Department of Electrical Engineering and Computer Science \\
  York University, Toronto, ON, Canada \\
  \texttt{\{amirmhd, hinat\}@yorku.ca}
}
\date{}
\begin{document}

\maketitle


\begin{abstract}
Wi-Fi–based human activity recognition (HAR) has emerged as a promising approach for contactless sensing, leveraging channel state information (CSI) collected from wireless transceivers.
While existing studies have primarily concentrated on single-user scenarios, real-world deployments often involve multi-user settings where concurrent users' movements induce overlapping CSI patterns that challenge conventional classification methods. To address this limitation, this paper introduces an attention-based multi-user activity recognition (AMAR) framework that formulates HAR as a set prediction problem.
The transformer-based architecture in AMAR leverages learnable query embeddings acting as specialized activity detectors, enabling the simultaneous identification of multiple activities from composite CSI representations.
{Moreover, to address deployment constraints, AMAR is designed in an edge-cloud split architecture form where lightweight convolutional networks on edge devices perform initial feature extraction, followed by residual vector quantization that achieves substantial bandwidth reduction while preserving activity-discriminative information. The cloud component performs final activity prediction through attention-based set matching, enabling the system to handle varying occupancy levels. Across classroom, meeting-room, and empty-room environments, on average AMAR nearly doubles the rate of perfectly predicting all concurrent activities compared to the best baseline. Moreover, it achieves an $F_1$-score of 53.4\% compared to 45.6\% 
for the best benchmark, and reduces occupancy estimation error by 74\%, while 
minimizing bandwidth substantially.}


\end{abstract}

\section{Introduction}
{Wireless sensing is a key enabling technique for emerging network applications including smart homes, healthcare facilities, and security systems. Traditional sensing relies on dedicated sensors such as cameras, motion detectors, or wearable devices. However, these solutions suffer from privacy concerns, installation costs, maintenance requirements, and network coverage gaps. Wi-Fi sensing offers an alternative that leverage the existing wireless communication infrastructure deployed in indoor environments. {Since channel state information (CSI) collected from Wi-Fi devices captures subtle variations in wireless signals as they interact with people and surrounding objects, it enables a range of applications such as human activity recognition (HAR), fall detection, crowd counting, and localization~\cite{TNNLS_activity, zou2018devicefree, spotfi}. Importantly, these applications operate in a privacy-preserving manner, as they do not require explicit disclosure of users’ identities.}

 While substantial progress has been made in the wireless sensing domain, there remains a considerable gap between laboratory demonstrations and real-world implementations \cite{tan2022survey, ma2019survey}. The reason  is that most of the existing Wi-Fi sensing research focuses on  environments where only a single user is present \cite{yousefi2017survey, borna}.  
 The core challenge with multi-user scenario is that when multiple people are present, their combined influence on the wireless channel cannot be easily attributed to individual users. One user's movements directly affect how another user's activities appear in the CSI measurements. Thus, predictions for different users must be interdependent rather than independent. This signal entanglement affects all sensing tasks—from accurately counting occupants and determining their locations, to recognizing individual activities and monitoring physiological signals \cite{tan2019multitrack}.  

Beyond signal entanglement, another fundamental challenge is the implicit coupling between multi-user HAR and occupancy count, resulting in a variable-size prediction problem, i.e., to determine both how many people are present and what each person is doing jointly. Existing works solved this specific challenge by first estimating the number of occupants, then decompose the mixed CSI signal into individual components, and finally classify each person's activity\cite{duan_wisdom_iotJ_2023, wang_tensorbeat_2017, he_IMar_acm_2022}.} However,  error in occupancy estimation propagates through signal decomposition and into HAR. Moreover, these methods {often} rely heavily on manual signal-processing  and feature engineering that fail to capture the complex, nonlinear interactions between users’ wireless signal patterns.

{Another line of research works handle unknown occupancy by coupling HAR with  an auxiliary task (e.g., localization \cite{MultiSenseX} or identity recognition \cite{wimans, multiUser_light_iot}) using deep learning (DL) techniques.}
While the existing DL solutions {have taken one step forward from classical signal processing and} captured complex patterns, their coupling with auxiliary tasks introduces limitations. Methods requiring location annotations demand additional labeled data that may not always be available, whereas those predicting user identity–activity pairs inadvertently learn user-specific characteristics rather than generalizable activity patterns, reducing their ability to handle new users. 

Beyond the aforementioned challenges, practical implementation of multi-user HAR networks suffers from resource constraints. In fact, edge devices in standard Wi-Fi routers have limited computational capacity, while transmitting raw CSI data to cloud servers consumes prohibitive transmission bandwidth~\cite{EfficientFi}. Any practical solution thus must address both the \textit{algorithmic challenge of multi-user HAR} and the \textit{system-level constraints of real-world deployment}\cite{multiUser_light_iot}.

{To address the aforementioned limitations, we develop a novel light-weight \textbf{A}ttention-based \textbf{M}ulti-user \textbf{A}ctivity \textbf{R}ecognition framework (\textbf{AMAR})\footnote{Code available at: \url{https://github.com/amirhosseinmhd/AMAR}}.  that can capture the interdependent nature of multi-user CSI patterns. Our approach requires no prior knowledge of occupancy levels, no signal decomposition, and no specialized hardware setup, making it deployable within any standard wireless sensing configuration. We formulate the multi-user HAR as a \textit{set prediction problem} which is then solved through a transformer-based architecture with \textit{learnable query embeddings}. Our approach naturally captures interdependent activities through attention mechanisms while achieving deployment feasibility via an edge-cloud split design with \textit{Residual Vector Quantization (RVQ)}. } To this end, our key contributions can be listed as follows:

{\begin{enumerate}
    \item We formulate multi-user HAR as a set prediction problem, fundamentally different from the existing works that require auxiliary user identity labels or location annotations. {To the best of our knowledge, this is the first end-to-end DL method that directly detects multiple concurrent users' activities through fully learnable neural network modules, solving the problem in its original form without coupling it to auxiliary tasks.} Our formulation naturally accommodates various occupancy levels through unordered set representation that also handles the permutation-invariant nature of the concurrent activities. 
    
    \item  We introduce AMAR, a transformer-based architecture that employs learnable query embeddings as specialized "activity detectors" to generate interdependent predictions from overlapping CSI patterns. Each query learns to focus on specific activity signatures through cross-attention with encoded CSI representations, while query-to-query self-attention enables coordination between predictions—essential for handling the complex multi-user CSI interactions. This design addresses the fundamental limitation of existing methods that treat each activity prediction independently, thus overlooking the coupled nature of multi-user CSI patterns.
    
    \item We propose an edge-cloud split architecture where edge devices perform initial compression and RVQ, while cloud servers handle transformer-based prediction. The RVQ layer discretizes continuous features into transmittable tokens, achieving over 99.2\% bandwidth reduction with minimal performance loss. Our complete system comprises only 0.32M parameters, enabling real-time inference on resource-constrained edge devices while maintaining communication efficiency between edge and cloud components.
    
    \item  { We introduce a comprehensive evaluation framework with appropriate metrics for multi-user sensing (i.e., count-based precision/recall, perfect prediction score (PPS), occupancy counting error (OCE)) and conduct extensive ablation studies across diverse indoor environments, such as classroom, meeting-room, and empty-room environments. Across all metrics, our proposed approach AMAR outperforms the state-of-the-art baselines. In addition, we demonstrate that AMAR achieves a 1.72× improvement in PPS and a 74\% reduction in OCE compared to state-of-the-art methods. }
\end{enumerate}}

{The remaining paper is structured as follows: Section~\ref{sec:background} provides background work on multi-user HAR. Section~\ref{sec:problem_formulation} defines the problem in terms of sensory input and target variables, and formulates the overall objective. Section~\ref{sec:architecture} details AMAR, our proposed edge-cloud architecture. Section~\ref{sec:evaluation_setup} introduces the experimental evaluation setup. Section~\ref{sec:numerical_results} presents numerical results comparing model performance, followed by ablation studies analyzing the impact of key design choices. Finally, Section~\ref{sec:conclusion} concludes the paper.}

\section{Background Work}
\label{sec:background}
{ The problem of multi-user HAR has been addressed to date by a variety of research works  leveraging tools from either DL or signal processing theory. In the following subsections, we provide a comprehensive review of the existing works.

\subsection{Signal Processing Approaches}
To date, several works decomposed the mixed CSI signals into individual user components before performing activity classification, using techniques like tensor decomposition \cite{wang_tensorbeat_2017, he_IMar_acm_2022} or blind source separation \cite{duan_wisdom_iotJ_2023}. These decomposition strategies faced some limitations. First, {they either assume that the number of occupants is given \cite{he_IMar_acm_2022, wang_tensorbeat_2017} or estimate it \cite{duan_wisdom_iotJ_2023} which remains error-prone}.} 
 Second, the assumption that multi-user CSI can be accurately separated into individual components oversimplified the complex, nonlinear interactions between multiple users' wireless signal patterns. These separation errors then propagate, thus  degrading the final classification performance.
 {More specifically, Wi-Multi \cite{wiMulti} estimated the number of users first and then applied handcrafted features followed by classical machine learning algorithm to classify concurrent activities—notably without any signal decomposition. Moreover, Wi-Multi predicts an unordered set of activities rather than assigning identities. However, since it relies on hand-engineered features, it struggled to automatically learn discriminative representations and to model interdependencies between simultaneous activities.}

 Other studies handled the problem by applying stronger priors. For instance, MultiSense \cite{zeng_multisense_2020} recovered multi-person respiration by separating periodic breathing components,  Wi-Run \cite{WIrun} estimated each runner’s steps from CSI dynamics, and \cite{Gait_yasaman} targeted multi-person gait identification behind a wall. These systems tackled specific periodic or biometric signals rather than general multi-user HAR.  WiMU \cite{WiMU} approached simultaneous gestures as a combinatorial synthesis problem and after detecting user count, it generated ``virtual" CSI by algebraically adding single-user gesture traces for every plausible combination, then classified the observed signal through nearest-sample matching.

MultiTrack \cite{tan2019multitrack} exploited multi-link diversity and combined multiple 5~GHz Wi-Fi channels into a single high-resolution profile to separate per-user reflections, reconstruct single-user signal views to suppress cross-interference, and then perform per-user localization and template-based HAR. Also, MUSE-Fi \cite{muse-fi} introduced proximity-based separation techniques that exploit near-field Wi-Fi variations to achieve physical user separation. However, the approach required each user to carry a Wi-Fi-enabled device within 0.2~m for activity classification, which limits practical deployment as users may not have such devices readily available.

{In summary, the aforementioned works—whether relying on signal decomposition, hand-crafted feature extraction, or hardware-tailored solutions—shared fundamental limitations that hinder their generalizability. Signal processing techniques typically require domain-specific assumptions about signal characteristics (e.g., periodicity, linearity of superposition) that may not hold in complex, real-world multi-user scenarios. Heuristic feature extraction methods depend  on manual engineering and domain expertise to identify discriminative features, making them sensitive to variations in environmental conditions, user behaviors, or activity types not anticipated during design. Similarly, hardware-based solutions  require specialized antenna configurations, multiple transceivers, or proximity-based devices that limit deployment flexibility. 

\subsection{Deep Learning Solutions}

{End-to-end DL methods}  can automatically learn discriminative representations directly from raw CSI data while jointly modeling the complex interdependencies between concurrent user activities.
{These methods, however, require fixed-size outputs—a mismatch when the number of active users is unknown. To resolve this, existing approaches typically perform HAR with an auxiliary task that predicts unknown occupancy variable and converts the problem into a fixed-dimensional prediction problem.}
Such approaches classify multi-user activities from mixed CSI signals without requiring explicit separation. For instance, WiMANS\cite{wimans} employed multi-label classification with Binary Cross Entropy (BCE) loss, treating each user-activity pair as an independent prediction task \cite{wimans, multiUser_light_iot}. More specifically, for the loss function, they introduced a prediction matrix where rows represent user identities and columns represent activity indices. Each element of this matrix predicts the activity of a predefined user. Later, the model is penalized using independent BCE loss function, essentially assuming one user can perform multiple activities. This causes the model to encode user-specific characteristics rather than generalizable activity features, limiting the ability to handle the combinatorial explosion of class labels   in multi-user HAR.

More recently, MultiSenseX~\cite{MultiSenseX} used a two-stage approach that first localizes users and then classifies activities at each detected location. However, the framework faced two major challenges. First, MultiSenseX required both location and activity annotations. This might not always be available as annotations are hard to acquire, especially when both types are needed for each sample. Second, when they predicted activities for each identified location, they treated the activity prediction task for each location independently and did not consider interdependent predictions. 

\subsection{Light-Weight Deep Learning Solutions}
{On a different note,  the deployment of DL models on the network edge devices must account for limited memory and bandwidth \cite{EfficientFi}. To address these constraints, recently BLTHAT \cite{multiUser_light_iot} proposed a lightweight design that performs multi-user HAR entirely on the edge at low computational cost. 
{In addition, RSCNet \cite{borna_rsc} compressed CSI on the edge, transmitted it to the cloud for reconstruction, and performed single-user HAR. Moreover, EfficientFi \cite{EfficientFi} not only compressed the edge representation, but also quantized it to further reduce communication overhead, thereby saving both computation and bandwidth. In our edge–cloud split architecture, we mitigate quantization error by introducing RVQ layers. }

\section{Multi-User HAR: A Set Prediction Approach}
\label{sec:problem_formulation}
{In this section, we formalize the multi-user HAR problem. We begin by specifying the sensory input and its signal model, clarifying how CSI sequences are represented. We then define the prediction target as an unordered set of activities with variable cardinality, reflecting an unknown number of occupants. With input and target in place, we cast the task as learning a mapping from CSI to sets and derive the optimization objective that trains this mapping. }

\subsection{CSI Measurement Model}
In CSI-based sensing, the wireless channel reflects environmental characteristics, enabling human activity detection from raw CSI measurements. When multiple users perform activities simultaneously, their combined influence on the multipath wireless channel creates distinct patterns that can be exploited for recognition. For a MIMO-OFDM wireless system with $\numtxant$ transmit antennas, $\numrxant$ receive antennas, and $\nrsubcarriers$ subcarriers, the CSI measurements at time $\timeidx$ and subcarrier $\subcarrieridx$ can be expressed as follows:
\begin{equation}
\notag
\featurevec(\timeidx,\subcarrieridx) = \sum_{p=1}^{P} \alpha_p(\timeidx)e^{-j2\pi f_\subcarrieridx \tau_p(\timeidx)},    
\end{equation}
where $\alpha_p(\timeidx)$ represents the complex attenuation of the $p$-th propagation path due to reflection and scattering, and $\tau_p(\timeidx)$ denotes the corresponding propagation delay. Over a time window of length $\initLength$, we collect CSI measurements as a complex-valued tensor $\featurevec \in \mathbb{C}^{T \times \numrxant \times \numtxant \times \nrsubcarriers}$.

\subsection{Multi-User HAR as Set Prediction}

Multi-user HAR presents a fundamental challenge: multiple individuals perform activities simultaneously, and there is no inherent ordering to these concurrent activities. A person walking while another sits has the same semantic meaning as a person sitting while another walks. This inherent permutation invariance suggests that the problem should naturally be formulated as a set prediction problem.

Previous DL approaches \cite{wimans, MultiSenseX} have attempted to handle this challenge by imposing priors—either user identity labels or spatial location annotations. However, these auxiliary constraints fundamentally misrepresent the problem's nature. We propose a novel formulation that models multi-user sensing directly as set prediction, respecting the problem's inherent permutation invariance. Our approach is inspired from a recent work in computer vision, where object detection has been reformulated as a set prediction problem because detected objects have no inherent ordering \cite{detr}.

Let $\mathcal{A} = \{1, 2, ..., \numActivities\}$ denotes the set of possible activities. For a given CSI measurement $\featurevec$, our objective is to predict the true activity set which is given by\footnote{We use the term "set" loosely throughout the paper, as our output can contain repeated elements when multiple users perform the same activity.}:
$$
\labelvec = \{\actionitem_1, \actionitem_2, ..., \actionitem_{\TrueNumPeople}\},
$$
{{where $\actionitem_i \in \mathcal{A}$ and $\TrueNumPeople$ denotes the true number of people inside the room which is unknown and varies across samples}.}
{An important consequence of this set-based formulation is that it naturally reduces the hypothesis space from $(\numActivities)^{\TrueNumPeople}$ ordered sequences to $\binom{\numActivities+{\TrueNumPeople}-1}{{\TrueNumPeople} }$ unordered multisets \cite{WiMU}. For example, with $\numActivities=9$ activities and up to ${\TrueNumPeople}=5$ concurrent users, this represents a 46-fold reduction in possible output configurations. While this computational benefit is significant, the primary contribution lies in correctly modeling the problem structure, i.e., by respecting the permutation invariance inherent in multi-user scenarios. Thus, our formulation enables the model to learn more generalizable activity patterns rather than spurious orderings or user-specific characteristics.}



\subsection{Optimization Objective}
\label{sec:opt_objective}

        


\begin{figure}[t]
    \centering
    \includegraphics[scale=0.6]{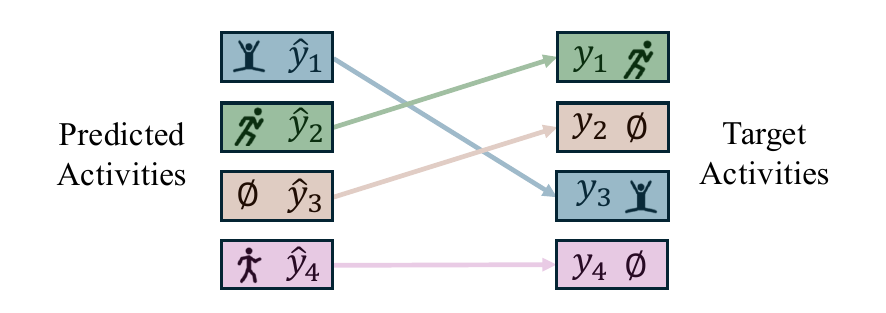}
    \caption{Bipartite matching between predictions and ground-truth targets. Given predictions $\hat{\labelvec}=\{\hat{\actionitem}_1,\dots,\hat{\actionitem}_{\numqueries}\}$ and padded targets $\widetilde{\labelvec}=\{\widetilde{\actionitem}_1,\dots,\widetilde{\actionitem}_{\numqueries}\}$, we build a cost matrix using the classification loss $\mathcal{L}_{\mathrm{cls}}(\widetilde{\actionitem}_i,\hat{\actionitem}_j)$ and use the Hungarian algorithm to find the optimal assignment $\sigma\in\setofPermutations$ that minimizes the total cost (cf. Eq.~\eqref{eq:set_loss}). The $\emptyset$ symbol denotes the \textit{``no person''} class, which eliminates the need for count estimation.}
    \label{fig:bipartite_matching}
\end{figure}
{Prior to setting up the objective, we must align the cardinalities of the ground-truth and predicted sets. To do so, we fix the model's output size to $\numqueries$ predictions, where $\numqueries$ is a design choice representing the system’s maximum supported occupancy. 
We introduce a special \textit{``no person'' class} $\emptyset$, so the model can output $\emptyset$ for positions corresponding to the absent users.
Each ground-truth set $\labelvec$ is padded with $\emptyset$ to expand its cardinality from $\TrueNumPeople$ to $\numqueries$, matching the prediction size; we refer to the padded ground truth as $\widetilde{\labelvec}$. 
This approach eliminates the need for explicit occupancy estimation as a separate task by implicitly asking the model to predict the \textit{``no person''} class, if appropriate.}
Formally,
\begin{equation}
\label{eq:padded_y}
\widetilde{\labelvec}=\{\actionitem_1,\dots,\actionitem_{\TrueNumPeople},\underbrace{\emptyset,\dots,\emptyset}_{\numqueries-\TrueNumPeople}\}.    
\end{equation}

{The set-prediction formulation requires a loss that is invariant to the ordering of predictions. We achieve this via optimal bipartite matching between predicted and ground-truth sets. Given the predicted set $\hat{\labelvec}=\{\hat{\actionitem}_1,\dots,\hat{\actionitem}_{\numqueries}\}$, where each $\hat{\actionitem}_i \in \mathbb{R}^{\numActivities+1}$ is a probability distribution over activities, and the padded ground-truth set $\widetilde{\labelvec}$, we define the \textit{matching loss} as:
\begin{equation}
\label{eq:set_loss}
\mathcal{L}(\widetilde{\labelvec},\hat{\labelvec})
\;=\;
\min_{\sigma \in \setofPermutations}
\;\sum_{i=1}^{\numqueries}
\mathcal{L}_{\mathrm{cls}}\!\big(\widetilde{\actionitem}_i,\;\hat{\actionitem}_{\sigma(i)}\big).
\end{equation}
{Here, ${\setofPermutations}$ denotes the set of all permutations of the prediction set $\hat{\labelvec}$, and $\mathcal{L}_{\mathrm{cls}}(\cdot,\cdot)$ is the cross-entropy loss between a ground-truth class label and a predicted probability distribution. The term $\hat{\actionitem}_{\sigma(i)}$ represents the prediction that permutation $\sigma$ assigns to match with ground-truth element $\widetilde{\actionitem}_i$. Since the predicted set is unordered, the optimization finds the permutation $\sigma$ that minimizes the total classification loss, ensuring that each prediction is matched with its most appropriate target regardless of the order in which predictions are generated.
}}

{We solve this bipartite matching problem using the Hungarian algorithm~\cite{kuhn1955hungarian}, which efficiently finds the optimal assignment in $O(\numqueries^3)$ time. The algorithm operates by constructing a cost matrix where each element represents the classification loss between a predicted activity and a target activity, then systematically finds the assignment that minimizes the total cost across all pairs.} 

In what follows, we design an architecture capable of (1) extracting discriminative features from overlapping CSI patterns, (2) generating $\numqueries$ distinct yet interdependent predictions that naturally form a set, and (3) operating efficiently within edge–cloud deployment constraints. 

\section{AMAR: Proposed Edge-Cloud Multi-User Activity Recognition Framework}}
\label{sec:architecture}
{AMAR processes multi-user CSI data through a three-stage pipeline designed to balance sensing accuracy with deployment constraints. The system splits processing between edge devices and cloud infrastructure to minimize computational overhead. First, CSI data undergoes \textbf{(1)} compression on edge devices, \textbf{(2)} then the quantization is performed at the edge and the quantized representations are transmitted to cloud servers followed by \textbf{(3)} transformer-based processing and final activity prediction. This edge-cloud split addresses both computational limitations of the edge devices and bandwidth constraints of  wireless communication networks. }

{The edge processing consists of a backbone network followed by RVQ. The backbone network compresses high-dimensional, noisy CSI data into dense feature representations using {\textit{depthwise separable convolutions (DSC)} \cite{chollet2017xception} and \textit{atrous (dilated) convolutions (AC)} \cite{yu2016multi}. }This compression reduces both temporal and channel dimensions, while filtering noise and preserving activity-relevant patterns. The RVQ layer then discretizes these features into tokens, reducing communication bandwidth  with minimal performance loss.}

{The cloud processing uses a transformer encoder-decoder architecture to generate final activity predictions. The transformer encoder captures long-range temporal dependencies in the quantized CSI representations through multi-head self-attention mechanisms, enabling the model to understand how CSI patterns evolve as multiple users perform various activities simultaneously. The decoder transforms these temporal representations into activity predictions using $\numqueries$ learnable query embeddings that act as specialized ``activity detectors.'' Each query learns through backpropagation to extract activity-relevant representations from the overlapping CSI patterns. The queries then interact via self-attention to model interdependencies between predicted activities, which is essential since CSI patterns from multiple users influence each other. Finally, feed-forward networks map each query's representation to activity predictions, and we train the entire system end-to-end using our matching loss function (see \Cref{eq:set_loss}) to handle the unordered nature of multi-user activity sets.}

\begin{figure}[t]
    \centering
    \includegraphics[width=0.5\textwidth]{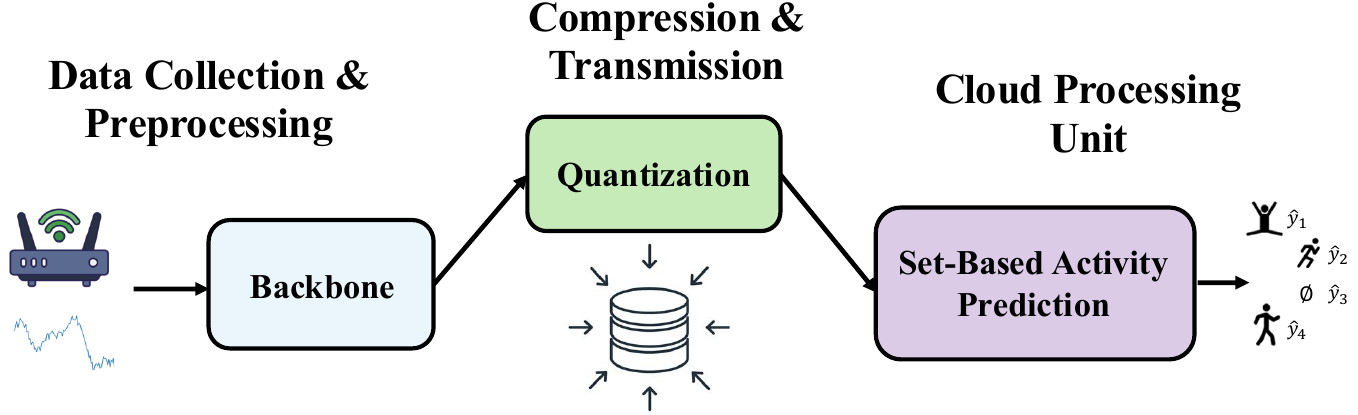}
\caption{Proposed multi-user HAR framework using transformer-based set prediction. The system splits processing between edge devices (backbone and quantization) and cloud servers (HAR) to enable light-weight multi-user sensing.}    \label{fig:highlevel_arch}
\end{figure}

\begin{figure*}[t]
    \centering
\includegraphics[width=1\textwidth]{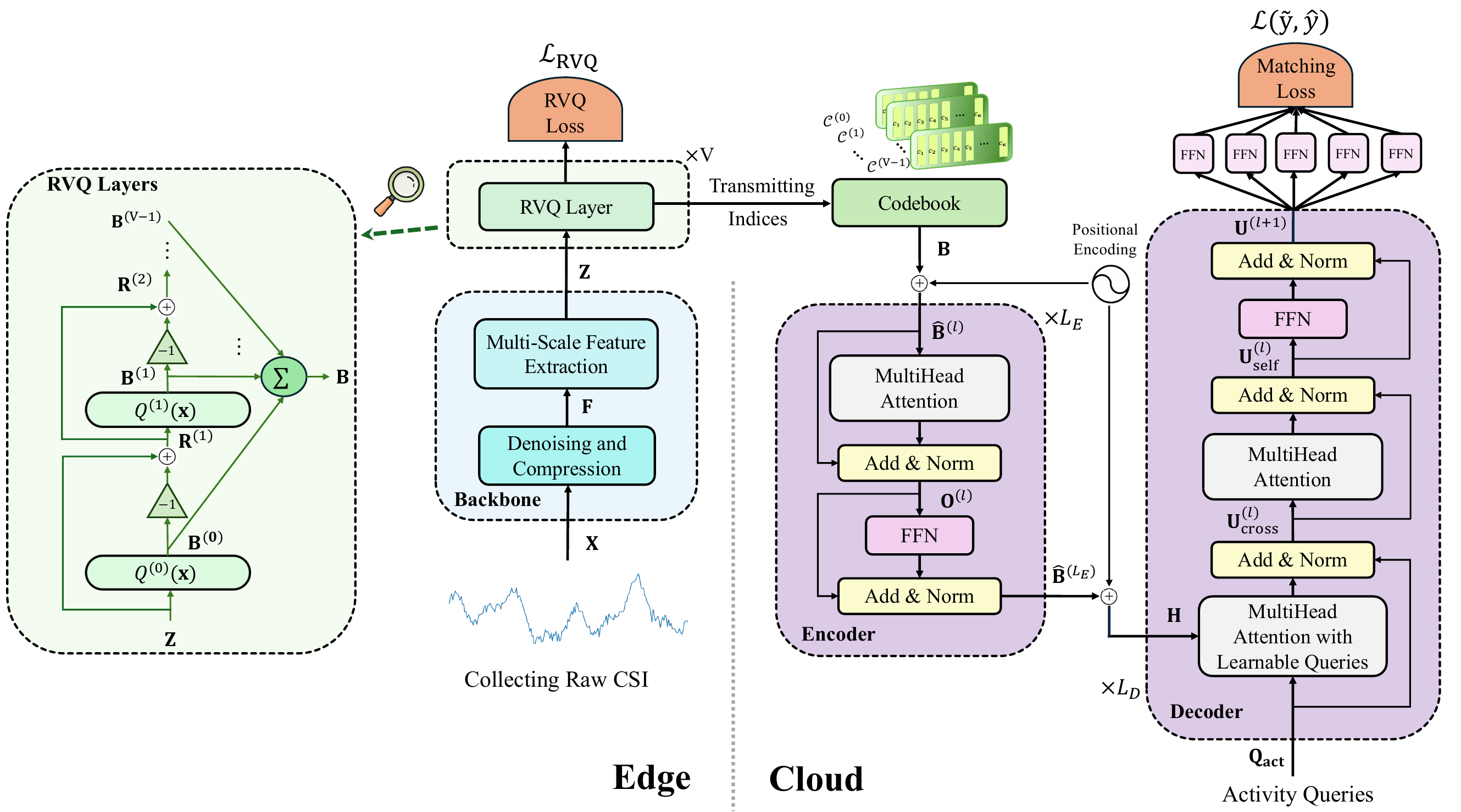}   
\caption{Detailed architecture showing edge-cloud processing split: backbone feature extraction and RVQ quantization on edge devices, followed by transformer encoder-decoder with learnable queries on cloud servers. The decoder generates multiple activity predictions using cross-attention and self-attention mechanisms. The system is trained end-to-end using \textit{the matching loss} (Eq.~\ref{eq:set_loss}) for set prediction and \textit{RVQ loss} (Eq.~\ref{eq:loss_RVQ}) for codebook learning.}
\label{fig:detailed_arch}
\end{figure*}

\subsection{Backbone-at-the-Edge} \label{sec:backbone}

The backbone operates on edge devices where computational resources are limited. This component serves three functions: (1) denoising and compressing high-dimensional CSI signals while preserving activity-specific patterns, (2) extracting multi-scale temporal features necessary for recognizing diverse activities, and (3) achieving substantial dimensionality reduction to enable efficient transformer processing. The third function is particularly important as transformers can overfit to noise patterns when fed raw, unfiltered signals like CSI \cite{THAT}, which significantly degrades recognition performance.

Our backbone architecture consists of two sequential blocks that progressively refine the CSI representations.
The first block referred to as \textit{Denoising and Compression Block} uses {DSC} to filter noise and compress the signal.
The second block referred to as \textit{Multi-scale Feature Extraction Block} then applies {AC} to {extract multi-scale and hierarchical temporal features from this compressed representation.}
Together, these blocks ensure that the backbone produces clean, compact, and information-rich features suitable for subsequent transformer-based processing on cloud servers.


\paragraph{\textbf{Denoising and Compression Block}}
Given an input CSI tensor, we first convert the complex-valued signal to its amplitude \cite{wimans,borna_rsc}.
We then reshape the amplitude  into a matrix representation $\mathbf{X} \in \mathbb{R}^{\initLength \times \initChannel}$, 
where $\initLength$ represents the initial temporal dimension and $\initChannel = \numrxant \times \numtxant \times \numsubcarriers$ represents the flattened channel dimensions. 
This flattening is justified by the statistical correlation observed across subcarriers, transmit antenna, and receive antenna dimensions, 
allowing us to treat them collectively as channels for our temporal analysis~\cite{wimans}.

{The first stage of our backbone focuses on noise filtering and compression through a series of DSC layers. 
The $i$-th DSC layer represented as $\text{DSC}^{(i)}(\mathbf{F}^{(i-1)}, \kernelSizeDSC{i}, \strideParam{i},\numOutputChannelsDSC{i})$ takes as input the previous layer's output $\mathbf{F}^{(i-1)}$
and the DSC layer hyperparameters, i.e., kernel size $\kernelSizeDSC{i}$, stride $\strideParam{i}$ and the output channels $\numOutputChannelsDSC{i}$.
A DSC layer executes  two operations:
\textbf{(1)} a depthwise convolution that applies independent 1D temporal convolutions to each input channel $c$ at time index $t$, producing intermediate features given by 
${d}_{t,c} = \sum_{j=0}^{\kernelSizeDSC{i}-1} w^{(i)}_{c,j} \cdot f^{(i-1)}_{t \cdot \strideParam{i} + j, c}$
where $w^{(i)}_{c,j}$ are learnable filter weights; and
\textbf{(2)} a point-wise convolution that mixes information across channels to produce the final output given by 
${j}_{t,m} = \sum_{c=1}^{\numOutputChannelsDSC{i-1}} v^{(i)}_{m,c} \cdot d_{t,c}$
where $v^{(i)}_{m,c}$ are learnable mixing weights.

We then stack $L_\text{DSC}$ DSC layers, each followed by BatchNorm and ReLU non-linearity given by:
\begin{align}
\notag
\mathbf{J}^{(i)} &= \text{DSC}^{(i)}(\mathbf{F}^{(i-1)}, \kernelSizeDSC{i}, \strideParam{i}, \numOutputChannelsDSC{i}), \\
\notag
\mathbf{F}^{(i)} &= \text{ReLU}(\text{BatchNorm}(\mathbf{J}^{(i)})),
\end{align}
where $\mathbf{F}^{(0)} = \mathbf{X}$ and we define $\mathbf{F} = \mathbf{F}^{(L_\text{DSC})}$. We choose $\strideParam{i}$ and $\numOutputChannelsDSC{i}$ to progressively compress the signal while preserving activity-related information.} The computational efficiency of DSC is particularly important given our resource constraints, as they reduce the parameter count from {$O(\kernelSizeDSC{i} \cdot \numOutputChannelsDSC{i-1} \cdot \numOutputChannelsDSC{i})$ to $O(\kernelSizeDSC{i} \cdot \numOutputChannelsDSC{i-1} + \numOutputChannelsDSC{i-1} \cdot \numOutputChannelsDSC{i})$} compared to standard convolutions.

\paragraph{\textbf{Multi-Scale Feature Extraction Block}}
{After initial compression, we employ sequential AC layers to capture multi-scale temporal dependencies. 

The $i$-th AC layer represented as $\text{AC}^{(i)}(\mathbf{A}^{(i-1)}, \kernelSizeAC, \dilationRate{i},\numOutputChannelsAC{i})$, takes as input the previous layer's output $\mathbf{A}^{(i-1)}$
and the AC layer hyperparameters, i.e., kernel size $\kernelSizeAC$, dilation rate $\dilationRate{i}$ and the output channels $\numOutputChannelsAC{i}$. Each  layer applies dilated convolution defined as  $y_{t,m} = \sum_{j=1}^{\kernelSizeAC} \sum_{c=1}^{\numOutputChannelsDSC{i-1}} w^{(i)}_{m,c,j} \cdot a^{(i-1)}_{t + j \cdot \dilationRate{i}, c}$. We implement a series of AC layers with exponentially increasing dilation rates, followed by BatchNorm and ReLU non-linearity:

\begin{align}
\notag
\mathbf{Y}^{(i)} &= \text{AC}^{(i)}(\mathbf{A}^{(i-1)}, \kernelSizeAC, \dilationRate{i}=2^{i-1}, \numOutputChannelsAC{i}), \\
\notag
\mathbf{A}^{(i)} &= \text{ReLU}(\text{BatchNorm}(\mathbf{Y}^{(i)})),
\end{align}}
{where $\mathbf{A}^{(0)} = \mathbf{F}$. Following the AC layers, we apply a final edge-side DSC layer $\text{DSC}^{(f)}$  for temporal downsampling and channel adjustment:
\begin{equation}
\notag
\mathbf{Z} = \text{DSC}^{{(f)}}(\mathbf{A}^{(L_\text{AC})}, k^{{(f)}}, s^{{(f)}}, \embedDim),
\end{equation}
where $L_\text{AC}$ denotes the number of AC layers, and the output $\mathbf{Z} \in \mathbb{R}^{\replength \times \embedDim}$ represents the compressed backbone features with temporal dimension $\replength$ and channel dimension $\embedDim$. This final layer serves as a transition block, compressing the multi-scale features into a compact representation  for subsequent quantization and transformer processing while preserving the rich temporal patterns captured by the AC layers.}

\subsection{RVQ Layers-at-the-Edge}
\label{sec:RVQ}
To utilize resources more efficiently, we use RVQ layers that further reduce communication costs. The idea is to send integer indices from edge devices instead of high-dimensional vectors, dramatically reducing communication overhead.

The approach works as follows: both edge devices and cloud servers maintain synchronized codebooks containing learned representative vectors called \textit{prototypes}. The edge device quantizes the backbone's output by finding the nearest prototype in its codebook and transmits only the corresponding index. The cloud server then reconstructs the prototype from its synchronized codebook \cite{EfficientFi}. However, single-layer quantization introduces substantial errors. We therefore adapt RVQ inspired by recent works in audio compression \cite{RVQ_audio} and human motion generation \cite{momask}, which progressively quantizes feature representations across multiple layers to preserve information while minimizing transmission bandwidth.

\paragraph{Quantization Process}
RVQ maintains $\numRVQLayers$ learned codebooks $\{\mathcal{C}^{(v)}\}_{v=0}^{\numRVQLayers-1}$, where each codebook $\mathcal{C}^{(v)} = \{c_i^{(v)}\}_{i=1}^{\codebookSize} \subset \mathbb{R}^{\embedDim}$ contains $\codebookSize$ prototype vectors.
For input feature matrix $\mathbf{Z} \in \mathbb{R}^{\replength \times \embedDim}$ from the backbone, we apply quantization independently to each temporal position $t$, where $\mathbf{z}_t$ denotes the $t$-th row of $\mathbf{Z}$.
The quantization at layer $v$ finds the nearest prototype, i.e.,
\begin{equation}
\label{eq:quantization_per_layer}
Q^{(v)}(\mathbf{x}) = \arg\min_{c \in \mathcal{C}^{(v)}} \|\mathbf{x} - c\|^2_2.
\end{equation}
The RVQ process begins by quantizing the input features:
\begin{equation}
\notag
\mathbf{b}^{(0)}_t = Q^{(0)}(\mathbf{z}_t), \quad \forall t \in \{1, \ldots, \replength\}.
\end{equation}
RVQ then computes the residual error and passes it to subsequent layers. Each layer $v$ quantizes the residual from the previous layer, i.e.,
\begin{equation}
\label{eq:flow_RVQ}
\mathbf{b}^{(v)}_t = Q^{(v)}(\mathbf{r}^{(v)}_t), \quad \mathbf{r}^{(v+1)}_t = \mathbf{r}^{(v)}_t - \mathbf{b}^{(v)}_t,
\end{equation}
where $\mathbf{r}^{(0)}_t = \mathbf{z}_t$. Here, $\mathbf{b}^{(v)}_t$ denotes the selected prototype from codebook $\mathcal{C}^{(v)}$, and we define $\gamma_t^{(v)}$ as the index of this prototype within the codebook, such that $\mathbf{b}^{(v)}_t = c_{\gamma_t^{(v)}}^{(v)}$. The final quantized representation sums all layer contributions:
\begin{equation}
\notag
\mathbf{b}_t = \sum_{v=0}^{\numRVQLayers-1} \mathbf{b}^{(v)}_t.
\end{equation}

Collecting all temporal positions yields the quantized matrix $\mathbf{B} \in \mathbb{R}^{\replength \times \embedDim}$ whose $t$-th row is $\mathbf{b}_t$. This progressive refinement captures both coarse activity patterns in early layers and fine-grained details in later layers. 

\paragraph{Training the Codebooks}
Unlike previous methods that optimize for signal reconstruction \cite{EfficientFi, borna_rsc}, our RVQ focuses solely on preserving discriminative features for HAR. We train the codebooks using a weighted sum of commitment loss and codebook loss as follows:
\begin{equation}
\label{eq:loss_RVQ}
\mathcal{L}_\text{RVQ} = \sum_{v=0}^{\numRVQLayers-1} \sum_{t=1}^{\replength} \left[ \|\mathbf{r}^{(v)}_t - \text{sg}[\mathbf{b}^{(v)}_t]\|_2 + \commitmentCost \|\text{sg}[\mathbf{r}^{(v)}_t] - \mathbf{b}^{(v)}_t]\|_2 \right],
\end{equation}
where $\text{sg}[\cdot]$ denotes the stop-gradient operation and $\commitmentCost$ controls the commitment cost.

To improve generalization and prevent overfitting to specific quantization patterns, we apply layer-wise dropout during training. Each RVQ layer $v \in \{0, \ldots, \numRVQLayers-1\}$ is randomly disabled with probability $p$. When a layer is dropped, its contribution $\mathbf{b}^{(v)}_t$ is set to zero and the residual $\mathbf{r}^{(v)}_t$ passes directly to the next layer. This regularization encourages the model to distribute information across multiple layers rather than relying heavily on early layers.
\paragraph{Edge-to-Cloud Transmission}

At the deployment stage, the edge device transmits only the quantized indices to the cloud server. For each RVQ layer $v$, we collect the indices across all temporal positions into a sequence $\Gamma^{(v)} = [\gamma_1^{(v)}, \gamma_2^{(v)}, \ldots, \gamma_{\replength}^{(v)}] \in \{1,\ldots,\codebookSize\}^{\replength}$, where each index $\gamma_t^{(v)}$ requires $\log_2(\codebookSize)$ bits. The edge device transmits $\{\Gamma^{(v)}\}_{v=0}^{\numRVQLayers-1}$ to the cloud server, which reconstructs the quantized features using its synchronized codebooks as $\mathbf{b}_t = \sum_{v=0}^{\numRVQLayers-1} c_{\gamma_t^{(v)}}^{(v)}$ for each temporal position $t$. The cloud server then processes these reconstructed features through the transformer encoder-decoder for final activity prediction.

{The multi-layer quantization  enables exponential growth in representational capacity. That is, a single codebook with $\codebookSize$ prototypes can only represent $\codebookSize$ distinct patterns. On the other hand, having $\numRVQLayers$ layers create $\codebookSize^{\numRVQLayers}$ distinct combinations. 

For example, with $\codebookSize=16$ prototypes per layer and $\numRVQLayers=4$ layers, our system can represent $16^4 = 65,536$ distinct patterns while transmitting only 4 indices per time step. This exponential growth far exceeds single-layer quantization, which would require a codebook of 65,536 prototypes to achieve the same expressiveness.}
{It is noteworthy that, using a single large codebook introduces significant challenges. At inference time, finding the nearest prototype in a codebook of size $\codebookSize^{\numRVQLayers}$ requires $\codebookSize^{\numRVQLayers}$ distance computations, compared to only $\numRVQLayers \times \codebookSize$ computations when using $\numRVQLayers$ layers of small codebooks. For our example configuration, this translates to approximately 1000× reduction in computational cost (64 distance computations versus 65,536). Additionally, training large codebooks presents optimization challenges, as ensuring all prototypes are effectively utilized requires careful regularization to prevent codebook collapse, where many entries remain unused. By combining small codebooks with multiple quantization layers, RVQ achieves dramatic bandwidth reduction and computational efficiency while preserving the discriminative information needed for multi-user sensing.}
\subsection{Encoder-at-the-Cloud}

While CNNs excel at extracting local features and performing initial denoising, they have limited capacity to model long-range temporal dependencies that are crucial for understanding complex activity patterns. The backbone's output captures local temporal structures, but activities often involve coordinated movements spanning several seconds with non-local temporal relationships. To address this limitation, we employ a transformer encoder that can relate distant temporal elements through its attention mechanism.

Our transformer encoder processes the reconstructed feature representation $\mathbf{B} \in \mathbb{R}^{{\replength} \times {\embedDim}}$ from the RVQ layer. Since the attention mechanism is permutation-invariant, we first add sinusoidal positional encodings to inject temporal order information as shown below:
\begin{equation}
\notag
{\encoderInput = \mathbf{B} + \text{SinusoidalPositionEncoding}(\mathbf{B})}.
\end{equation}
These positional encodings enable the model to distinguish between identical patterns occurring at different time steps, allowing it to learn temporal relationships crucial for HAR. 
{The core of our transformer encoder is the multi-head attention mechanism. This mechanism operates on three inputs that produces queries ($\mathbf{Q}$), keys ($\mathbf{K}$), and values ($\mathbf{V}$) through learned linear projections. Given three potentially different inputs $\mathbf{X}_Q$, $\mathbf{X}_K$, and $\mathbf{X}_V$, we compute $\mathbf{Q} = \mathbf{X}_Q\mathbf{W}^Q$, $\mathbf{K} = \mathbf{X}_K\mathbf{W}^K$, and $\mathbf{V} = \mathbf{X}_V\mathbf{W}^V$. The attention operation computes how much each query position should attend to each key position by taking their dot product, scaling by $\sqrt{{\keyDim}}$ where {$\keyDim$} is the key dimension, and applying softmax to get attention weights that aggregate information from the values. Multi-head attention applies this mechanism {$\numAttentionHeads$} times in parallel with different learned projections, allowing the model to capture different types of relationships simultaneously. The outputs from all heads are concatenated and projected through a final linear transformation $\mathbf{W}^O$:}
\begin{equation}
\notag
\text{MultiHead}(\mathbf{X}_Q, \mathbf{X}_K, \mathbf{X}_V) = \text{Concat}(\text{head}_1, \ldots, \text{head}_{{\numAttentionHeads}})\mathbf{W}^O,
\end{equation}
where each head is computed as follows:
\begin{equation}
\notag
\text{head}_i = \text{softmax}\left(\frac{\mathbf{Q}_i \mathbf{K}_i^T}{\sqrt{{\keyDim}}}\right) \mathbf{V}_i.
\end{equation}
{{In the encoder, we use self-attention by setting all three inputs to be identical, i.e., $\mathbf{X}_Q = \mathbf{X}_K = \mathbf{X}_V = \encstate{l}$, where $\encstate{l}$ is the representation at layer $l$. This allows each temporal position to dynamically attend to all other positions in the sequence, capturing dependencies regardless of their distance. Each self-attention is followed by a feed-forward network, with residual connections and layer normalization after each sub-layer given as follows:}}
\begin{align}
\notag
&{\encattn{l}} {= \text{LayerNorm}(\encstate{l} + \text{MultiHead}(\encstate{l}, \encstate{l}, \encstate{l}))},\\
\notag
&{\encstate{l+1}} {= \text{LayerNorm}(\encattn{l} + \text{FFN}(\encattn{l}))},
\end{align}
{where $\encstate{0} = \encoderInput$ is the input with positional encodings and $\text{FFN}(\cdot)$ denotes a two-layer feed-forward network with ReLU activation.} The stacked architecture with {$\numEncoderLayers$} encoder layers progressively refines these temporal representations, with each layer incorporating information from across the entire sequence.
{After {$\numEncoderLayers$} layers, we obtain $\encstate{\numEncoderLayers} \in \mathbb{R}^{{\replength} \times {\embedDim}}$, which encodes rich temporal dependencies from the CSI data.} { Before passing these representations to the decoder, we re-apply the same sinusoidal positional encodings to obtain the final encoder output as follows:
\begin{equation}
\notag
{\encoutput = \encstate{\numEncoderLayers} + \text{SinusoidalPositionEncoding}(\encstate{\numEncoderLayers})}. \label{eq:output_encoder_position}
\end{equation}
This re-addition of positional information helps the decoder maintain awareness of the temporal order of the encoded features, which is essential for the cross-attention mechanism to properly align decoder queries with the correct temporal positions in the encoder output.}
This encoded representation then serves as the foundation for the decoder to generate activity predictions, providing both the keys and values for cross-attention operations in the subsequent decoder layers.

\subsection{Decoder-at-the-Cloud with Learnable Queries}

{The decoder faces a unique challenge: it must generate $\numqueries$ distinct activity predictions from the encoded CSI features $\encoutput$, where each prediction identifies one user's activity. However, these predictions cannot be independent. As established in Section~\ref{sec:opt_objective}, multi-user CSI patterns are inherently coupled, i.e.,  one user's movements directly affect how another user's activities appear in the wireless signal. Traditional approaches that apply independent classifiers to shared features fail to capture these interdependencies, limiting their ability to correctly interpret overlapping activity signatures.}

{We address this challenge through learnable query embeddings that serve as specialized activity detectors. Rather than deriving queries from the input data, we treat them as trainable parameters of the network: $\queryemb \in \mathbb{R}^{\numqueries \times {\embedDim}}$. Each query embedding learns through backpropagation to focus on specific activity patterns in the encoded CSI representations. First, these queries extract information independently through cross-attention with the encoder output, enabling each to specialize in detecting different activities. Second, the queries communicate through self-attention, coordinating their predictions to account for signal interdependencies. 

This design naturally handles varying occupancy—when fewer than $\numqueries$ users are present, some queries learn to predict "no person," effectively adapting to different occupancy levels without explicit counting.}
{The decoder processes these learnable queries through {$\numDecoderLayers$} stacked layers, where each layer refines the queries' understanding through three operations. For the $l$-th layer, the input consists of the refined representation from the previous layer $\decstate{l-1}$ combined with the learnable query embeddings.} The steps are detailed as follows:

\textbf{Step 1: Cross-Attention with Learnable Queries.} The learnable queries attend to the encoder output to gather activity-specific information:

\scalebox{0.9}{
\begin{minipage}{0.5\textwidth}
\begin{equation}
\notag
\decstatecross{l} = \text{LayerNorm}(\decstate{l-1} + \text{MultiHead}(\decstate{l-1} + \queryemb, \encoutput, \encoutput)),
\end{equation}
\end{minipage}
}
\vspace{10pt}

where $\decstate{0} = \mathbf{0} \in \mathbb{R}^{\numqueries \times {\embedDim}}$ and $\encoutput \in \mathbb{R}^{{\replength} \times {\embedDim}}$ is the encoder output. The learnable queries $\queryemb$ provide the query content, allowing each to scan the temporal sequence and focus on patterns matching its learned specialization.

\textbf{Step 2: Query-to-Query Self-Attention.} The queries interact to coordinate their predictions:

\scalebox{0.9}{\begin{minipage}{0.5\textwidth}
\begin{equation}
\notag
\decstateself{l} = \text{LayerNorm}(\decstatecross{l} + \text{MultiHead}(\decstatecross{l}, \decstatecross{l}, \decstatecross{l})),
\end{equation}\end{minipage}
}
\vspace{8pt}

 Since CSI patterns from multiple users overlap, one query's detection of "walking" influences another query's interpretation of ambiguous patterns, ensuring comprehensive activity coverage without duplicates.

\textbf{Step 3: Feed-Forward Refinement.} Each query undergoes non-linear transformation:
\begin{equation}
\notag
\decstate{l} = \text{LayerNorm}(\decstateself{l} + \text{FFN}(\decstateself{l})),
\end{equation}
where FFN refines each query's representation based on the gathered information.

\textbf{Step 4: Final Classification.} After each decoder layer $l$, we apply a classification head:
\begin{equation}
\notag
\hat{\labelvec}^{(l)} = \text{softmax}(\mathbf{W}_{{\mathrm{cls}}} \cdot \decstate{l} + \mathbf{b}_{{\mathrm{cls}}}),
\end{equation}
where $\hat{\labelvec}^{(l)} \in \mathbb{R}^{\numqueries \times (\numActivities+1)}$ contains each query's prediction over all activity classes plus "no person". During training, we apply deep supervision by computing auxiliary losses at intermediate decoder layers using matching indices from the final layer, weighted by $\auxLossCoef$, to accelerate convergence.

Through these {$\numDecoderLayers$} layers, the learnable queries progressively specialize. They evolve to become activity detectors that coordinate through self-attention to handle the fundamental challenge of recognizing multiple concurrent activities from overlapping CSI patterns.

\section{Evaluation Set-Up and Baselines}

\label{sec:evaluation_setup}

{This section  begin by describing the evaluation dataset and its key characteristics. Importantly, we introduce the metrics for correctly measuring the performance of multi-user HAR systems. We then present the baselines that represent the current state-of-the-art. Finally, we specify our experimental configuration and model hyperparameters.}

\begin{algorithm}[h]
\caption{AMAR: End-to-End Training and Inference}
\begin{algorithmic}[1]
\item[] \textbf{Input:} Training/validation/test data $D_{\text{train}}, D_{\text{val}}, D_{\text{test}}$; Hyperparameters: epochs $\numEpochs$, batch size $\batchSize$, learning rate $\learningRate$, queries $\numqueries$, RVQ layers $\numRVQLayers$, codebook size $\codebookSize$.
\STATE \textbf{Initialize:} Backbone (DSC + AC layers), RVQ codebooks $\{\mathcal{C}^{(v)}\}_{v=0}^{\numRVQLayers-1}$, transformer encoder ($\numEncoderLayers$ layers), decoder ($\numDecoderLayers$ layers) with learnable queries $\queryemb$
\STATE \textbf{Training Phase (Repeat for $\numEpochs$ epochs):}
    \FOR{each sample $\{\mathbf{X}, \labelvec\}$ in $D_{\text{train}}$}
        \STATE \textit{// Forward Pass}
        \STATE $\mathbf{Z} \leftarrow \text{Backbone}(\mathbf{X})$ \hfill $\triangleright$ Feature extraction
        \STATE $\{\Gamma^{(v)}\}_{v=0}^{\numRVQLayers-1} \leftarrow \text{RVQ}(\mathbf{Z})$ \hfill $\triangleright$ Obtain Indices
        \FOR{each $t \in \{1,...,\replength\}$:}
        \STATE $\mathbf{b}_{t} \leftarrow \sum_{v=0}^{\numRVQLayers-1} c_{\gamma^{(v)}_t}^{(v)}$
        \hfill $\triangleright$ Reconstruct row $t$ of $\mathbf{B}$
        \ENDFOR
        \STATE $\encoutput \leftarrow \text{Encoder}(\mathbf{B} + \text{PE})$ 
        \STATE $\hat{\labelvec} \leftarrow \text{Decoder}(\queryemb, \encoutput)$ \hfill $\triangleright$ Activity prediction
        \STATE \textit{// Loss Computation and Update}
        \STATE $\widetilde{\labelvec} \leftarrow \text{Pad}(\labelvec, \emptyset)$ \hfill $\triangleright$ Pad with ``no person'' class (Eq.~\ref{eq:padded_y})
        \STATE $\mathcal{L} \leftarrow \mathcal{L}(\widetilde{\labelvec} ,\hat{\labelvec}) + \mathcal{L}_{\text{RVQ}}$ \hfill $\triangleright$ Matching loss (Eq.~\ref{eq:set_loss}) + RVQ loss (Eq.~\ref{eq:loss_RVQ})
        \STATE Update parameters via backpropagation
    \ENDFOR
    \STATE Evaluate on $D_{\text{val}}$ and save best model
\STATE \textbf{Inference Phase:}
\FOR{each $\mathbf{X} \in D_{\text{test}}$}
    \STATE \textbf{Edge:} $\mathbf{Z} \leftarrow \text{Backbone}(\mathbf{X})$, quantize to obtain indices $\{\Gamma^{(v)}\}_{v=0}^{\numRVQLayers-1}$
    \STATE \textbf{Edge$\rightarrow$Cloud:} Transmit indices $\{\Gamma^{(v)}\}_{v=0}^{\numRVQLayers-1}$
    \STATE \textbf{Cloud:} Reconstruct $\mathbf{b}_t \leftarrow \sum_{v=0}^{\numRVQLayers-1} c_{\gamma^{(v)}_t}^{(v)}$, \\predict $\hat{\labelvec} \leftarrow \text{Decoder}(\queryemb, \text{Encoder}(\mathbf{B}))$
\ENDFOR
\STATE \textbf{Output:} Activity predictions $\{\hat{\labelvec}\}$, evaluation metrics (PPS, $F_1$-score, OCE)
\end{algorithmic}
\end{algorithm}

\subsection{Dataset Characteristics}

We evaluate our proposed approach using the WiMANS dataset \cite{wimans}, focusing on the 5 GHz Wi-Fi band. To our knowledge, WiMANS represents the only publicly available dataset specifically designed for multi-user HAR. Despite being the sole public dataset, WiMANS is comprehensive for rigorous evaluation due to its large scale and environmental diversity. The dataset captures nine different activities (pick up, stand up, walk, rotation, jump, wave, lie down, sit down, nothing) performed simultaneously by up to five users across three distinct indoor environments: classroom, meeting-room, and empty-room, totaling 5,643 samples. 
Unlike single-user datasets, WiMANS captures concurrent activities from multiple users without sampling bias, reflecting the true complexity of real-world multi-user scenarios. The dataset employs a MIMO system with three transmitting and three receiving antennas, operating at a sampling rate of 1000 Hz across 30 subcarriers. Each three-second sample yields a CSI tensor $\featurevec \in \mathbb{R}^{3000 \times 30 \times 3 \times 3}$, where 3000 represents the temporal dimension, 30 denotes the subcarriers, and the last two dimensions correspond to the transmit-receive antenna pairs.

\subsection{Evaluation Metrics}
Multi-user HAR requires specialized metrics beyond standard classification accuracy. Recent works \cite{wimans, MultiSenseX} use "accuracy" metrics based on user-identity pairs that fail to properly evaluate activity count predictions. We employ a  comprehensive evaluation framework to measure various aspects of model performance.
For fair comparison across methods with different output formats, we first standardize all predictions to a unified representation: $\predictedlabelvec \in \mathbb{N}^{\numActivities}$, where each element represents the count of users performing the corresponding activity. Methods predicting user-identity pairs or location-activity pairs \cite{wimans, MultiSenseX} are converted by aggregating predictions across user and location slots respectively. The key performance metrics are given below:

\subsubsection{Count-Based Precision, Recall, and $F_1$-Score}
Unlike binary classification where each prediction is simply correct or incorrect, our multi-user setting requires counting how many instances of each activity are correctly identified. For each  sample ${\sampleidx}$ and activity ${\activityidx}$, we define:
\begin{align}
\begin{aligned}
\text{TP}_{\sampleidx,\activityidx} &= \min(y_{\sampleidx,\activityidx}, \hat{y}_{\sampleidx,\activityidx}), \quad
\text{FP}_{\sampleidx,\activityidx} = \max(0, \hat{y}_{\sampleidx,\activityidx} - y_{\sampleidx,\activityidx}), \\
\text{FN}_{\sampleidx,\activityidx} &= \max(0, y_{\sampleidx,\activityidx} - \hat{y}_{\sampleidx,\activityidx}),
\notag
\end{aligned}
\end{align}
where TP counts correctly identified instances, FP counts over-predictions, and FN counts missed instances. For example, if in a sample 3 people are walking and the model predicts 2 people walking, we get $\text{TP}_{\sampleidx,\activityidx}=2$, $\text{FN}_{\sampleidx,\activityidx}=1$, $\text{FP}_{\sampleidx,\activityidx}=0$.
We compute the expected values across all samples to obtain per-activity metrics, i.e.,
\begin{equation*}
\text{TP}_{\activityidx} = \mathbb{E}_{\sampleidx}[\text{TP}_{\sampleidx,\activityidx}],\quad
\text{FP}_{\activityidx} = \mathbb{E}_{\sampleidx}[\text{FP}_{\sampleidx,\activityidx}],\quad
\text{FN}_{\activityidx} = \mathbb{E}_{\sampleidx}[\text{FN}_{\sampleidx,\activityidx}],
\end{equation*}
In practice, we estimate these expectations using sample averages, i.e., $\text{TP}_{{\activityidx}} = \frac{1}{\samplesize}\sum_{{\sampleidx=1}}^{{\samplesize}}\text{TP}_{{\sampleidx,\activityidx}}$.

For activity $\activityidx$ the precision and recall are then computed as:
\begin{align*}
\text{Precision}_{{\activityidx}} &= \frac{\text{TP}_{{\activityidx}}}{\text{TP}_{{\activityidx}} + \text{FP}_{{\activityidx}}}, \quad
\text{Recall}_{{\activityidx}} = \frac{\text{TP}_{{\activityidx}}}{\text{TP}_{{\activityidx}} + \text{FN}_{{\activityidx}}}, \\
F_{1,{\activityidx}} &= \frac{2 \cdot \text{Precision}_{{\activityidx}} \cdot \text{Recall}_{{\activityidx}}}{\text{Precision}_{{\activityidx}} + \text{Recall}_{{\activityidx}}}.
\end{align*}

When the denominator is zero (no predictions or ground truth for an activity), we define the metric as zero. Final metrics are macro-averaged across all activities:
\begin{equation}
\notag
\text{Precision} = \frac{1}{\numActivities}\sum_{\activityidx=1}^{\numActivities}\text{Precision}_{\activityidx}, \quad \!\!\!
\text{Recall} = \frac{1}{\numActivities}\sum_{\activityidx=1}^{\numActivities}\text{Recall}_{\activityidx}, 
\end{equation}
\begin{equation}
    F_1\text{-score} = \frac{1}{\numActivities}\sum_{\activityidx=1}^{\numActivities} F_{1,\activityidx}.
    \notag
\end{equation}
This macro-averaging treats all activities equally regardless of their frequency. This approach is appropriate when we consider detecting all activities with equal importance~\cite{powers2020evaluationprecisionrecallfmeasure}.

\subsubsection{Perfect Prediction Score (PPS)}
{We measure exact set prediction accuracy using PPS, which represents the probability that a model correctly predicts all activity counts, i.e.,}
\begin{equation}
\label{eq:perfect_prediction}
    \text{PPS} = \mathbb{E}_{{\sampleidx}} [\mathbb{I}(\labelvec_\sampleidx = \predictedlabelvec_\sampleidx)],
\end{equation}
where the expectation is taken over test samples. PPS is a stringent metric that requires perfect accuracy of the whole vector $\predictedlabelvec_\sampleidx$, i.e., accuracy across all activities simultaneously.

\subsubsection{Occupancy Counting Error (OCE)}
We evaluate models' ability to count the total number of people using mean absolute error, which computes the expected absolute difference between predicted and actual occupancy, i.e.,
\begin{equation}
    \text{OCE} = \mathbb{E}_{{\sampleidx}} \left[\left|\sum_{{\activityidx=1}}^{\numActivities} {y_{\sampleidx,\activityidx}} - \sum_{{\activityidx=1}}^{\numActivities} {\hat{y}_{\sampleidx,\activityidx}}\right|\right],
    \label{eq:MAE_count}
\end{equation}
where $\sum_{{\activityidx=1}}^{\numActivities} {y_{\sampleidx,\activityidx}}$ shows the count of people for the sample $\labelvec_\sampleidx$. This metric  estimate room occupancy which is fundamental for  multi-user sensing applications.

\subsection{Considered Baselines}
\label{sec:baselines}
We compare our approach against three types of end-to-end baselines. The first type uses BCE for multi-label classification, treating each user-activity pair independently. The second type uses Direct Error Minimization (DEM) through regression to predict activity counts directly. The third baseline, MultiSenseX, uses a two-stage approach that first localizes users then classifies their activities. For BCE and DEM approaches, we test two different encoders: THAT \cite{THAT} and ABLSTM \cite{elkelany2023wifi}. This gives us four baseline variants: BCE-THAT, BCE-ABLSTM, DEM-THAT, and DEM-ABLSTM.

\subsubsection{BCE-Based Baselines}
The BCE loss is adopted from WiMANS \cite{wimans} which formulates multi-user sensing as a multi-label classification problem. The model predicts probabilities for each user-activity pair independently using BCE loss, i.e.,
\begin{equation}
\label{eq:WiMANS_LOSS} \notag
    \loss{BCE}{\predictedlabelvec}{\labelvec} = -\sum_{j=1}^{54} \labelvec_j \cdot \log(\sigmoid(\predictedlabelvec_j)) + (1-\labelvec_j) \cdot \log(1-\sigmoid(\predictedlabelvec_j)).
\end{equation}
The approach flattens all labels and trains the model in a supervised fashion. Since this model can predict multiple activities for one user, we post-process the outputs by selecting the most probable activity for each user. We then aggregate these predictions into a 9-element activity vector that counts how many users perform each activity. We test this approach with two best-performing encoders: THAT \cite{THAT} and ABLSTM \cite{elkelany2023wifi}, which we refer to as BCE-THAT and BCE-ABLSTM.
\subsubsection{DEM-Based Baselines}

{We consider the DEM approach as an additional baseline to investigate whether direct activity count prediction, without explicitly modeling interdependencies between users, can achieve comparable performance to our set prediction framework. The approach uses regression to predict activity counts directly, where the model outputs a vector $\predictedlabelvec \in \realcoorspace{9}$ with each element representing the predicted count of people performing each activity. Unlike our method that captures interdependencies through attention mechanisms, DEM treats each activity count as an independent regression target. To handle the discrete nature of person counts while maintaining smooth gradients during training, we use the smooth} L1 loss function defined as follows \cite{girshick2015fastrcnn}:
\begin{equation} \notag
\loss{smooth}{\predictedlabelvec}{\labelvec} = 
\begin{cases}
0.5(\predictedlabelvec - \labelvec)^2/{\smoothLossBeta}, & \text{if } |\predictedlabelvec - \labelvec| < {\smoothLossBeta}, \\
|\predictedlabelvec - \labelvec| - 0.5{\smoothLossBeta}, & \text{otherwise},
\end{cases}
\end{equation}
where $\smoothLossBeta$ controls the transition between L2 and L1 behavior. Since the loss function outputs real numbers, we discretize and clip the output to obtain predictions within the valid label space. Like the BCE baselines, we test this approach with both ABLSTM and THAT encoders, referring to them as DEM-THAT and DEM-ABLSTM.
\subsubsection{MultiSenseX}
MultiSenseX \cite{MultiSenseX} uses a two-stage end-to-end approach. The method first localizes users within the environment, then classifies activities at each detected location. In the first stage, the model predicts user presence at five predefined spatial locations using multi-label classification. In the second stage, independent feed-forward layers predict activities at locations where users were detected. Each location-specific classifier operates independently without considering interdependencies between users' activities \cite{MultiSenseX}. For fair comparison, we convert MultiSenseX outputs to our standardized format by aggregating predicted activities across all locations into a 9-element activity count vector.

\subsection{Model Specifications}
\begin{table}[H]
\centering
\caption{Model Configuration Parameters}
\label{tab:model_specs}
\footnotesize
\begin{tabular}{>{\raggedright}p{4cm}>{\centering\arraybackslash}p{2.5cm}}
\toprule
\textbf{Parameter} & \textbf{Value} \\
\midrule
Number of queries & $\numqueries = 6$ \\
Attention heads & {$\numAttentionHeads = 4$} \\
Number encoder layers & {$\numEncoderLayers = 4$} \\
Number decoder layers & {$\numDecoderLayers = 6$} \\
Embedding dimension & {$\embedDim = 64$} \\
Embedding length & {$\replength = 188$} \\
\midrule
Number RVQ layers & {$\numRVQLayers = 4$} \\
Codebook size & {$\codebookSize = 16$} \\
Commitment cost & {$\commitmentCost = 0.5$} \\
Auxiliary loss coefficient & {$\auxLossCoef = 0.25$} \\
\midrule
Learning rate & {$\learningRate = 5 \times 10^{-4}$} \\
Epochs & {$\numEpochs = 300$} \\
Batch size & {$\batchSize = 16$} \\
Weight decay & {$\weightDecay = 1 \times 10^{-4}$} \\
\bottomrule
\end{tabular}
\end{table}

Our transformer-based architecture employs $\numEncoderLayers = 4$ encoder layers and $\numDecoderLayers = 6$ decoder layers, processing $\numqueries = 6$ learnable query embeddings to handle up to 6 concurrent users. The model uses $\numAttentionHeads = 4$ attention heads per layer with an embedding dimensionality of $\embedDim = 64$. {The backbone network consists of $L_\text{DSC} = 3$ depthwise separable convolutional layers followed by $L_\text{AC} = 3$ atrous convolutional layers, which compress the input CSI to a temporal dimension of ${\replength} = 188$ time steps.} For efficient edge-cloud communication, we implement $\numRVQLayers = 4$ RVQ layers with $\codebookSize = 16$ prototype vectors per codebook. The model is trained for $\numEpochs = 300$ epochs using the Adam optimizer with a learning rate of $\learningRate = 5 \times 10^{-4}$ and weight decay of $\weightDecay = 1 \times 10^{-4}$. \textbf{Table~\ref{tab:model_specs}} summarizes the model configuration parameters. The dataset is split into training ($D_{\text{train}}$, 80\%), validation ($D_{\text{val}}$, 10\%), and testing ($D_{\text{test}}$, 10\%) sets. We report results on $D_{\text{test}}$ using the model configuration that performed best on $D_{\text{val}}$ based on PPS across all approaches. All results are averaged over 8 random seeds to obtain a better estimate of performance.

\section{Numerical Results and Discussions}
\label{sec:numerical_results}
In this section, we compare the performance of the proposed framework AMAR with the various baselines mentioned in \textbf{Section~\ref{sec:baselines}} in  a variety of environments followed by complexity analysis of the proposed algorithm.

\subsection{Performance Comparison}

\subsubsection{Multi-User HAR}
\begin{table*}[h]
\caption{Performance Comparison Across Different Environments (All Values Are Shown in \%)}
\centering
\small
\renewcommand{\arraystretch}{1.5}
\label{tab:results}
\resizebox{\textwidth}{!}{%
\begin{tabular}{l|cccc|cccc|cccc}
\toprule
 & \multicolumn{4}{c|}{\textbf{Empty-room}} & \multicolumn{4}{c|}{\textbf{Classroom}} & \multicolumn{4}{c}{\textbf{Meeting-room}} \\
\textbf{Method} & \textbf{PPS} & \textbf{$F_1$} & \textbf{Prec.} & \textbf{Rec.} & \textbf{PPS} & \textbf{$F_1$} & \textbf{Prec.} & \textbf{Rec.} & \textbf{PPS} & \textbf{$F_1$} & \textbf{Prec.} & \textbf{Rec.} \\
\midrule
MultiSenseX & 8.99 & 40.24 & 47.13 & 39.78 & 8.13 & 37.50 & 42.38 & 37.20 & 11.13 & 38.84 & 45.19 & 38.74 \\
BCE-THAT & 19.76 & 44.78 & 51.98 & 39.42 & 20.11 & 36.53 & 48.36 & 30.26 & 19.11 & 33.71 & 42.70 & 28.86 \\
DEM-THAT & 17.78 & 50.74 & 56.30 & 46.24 & 17.06 & 43.34 & 54.76 & 37.24 & 16.99 & 42.62 & \underline{54.49} & 36.37 \\
BCE-ABLSTM & 16.07 & 38.50 & 44.42 & 35.67 & 17.79 & 35.88 & 41.35 & 32.53 & 14.88 & 32.17 & 35.91 & 30.38 \\
DEM-ABLSTM & 10.71 & 43.09 & 52.77 & 38.08 & 14.29 & 38.60 & 45.93 & 34.72 & 10.38 & 31.46 & 47.20 & 24.65 \\
\textbf{AMAR w/o RVQ} & \underline{35.98} & \underline{59.51} & \underline{59.77} & \underline{60.12} & \underline{37.63} & \underline{57.84} & \underline{58.11} & \underline{58.42} & \underline{32.21} & \underline{50.75} & \underline{\underline{51.24}} & \underline{51.95} \\
\textbf{AMAR} & \underline{\underline{34.58}} & \underline{\underline{58.45}} & \underline{\underline{58.76}} & \underline{\underline{58.93}} & \underline{\underline{37.03}} & \underline{\underline{56.57}} & \underline{\underline{57.05}} & \underline{\underline{56.91}} & \underline{\underline{29.89}} & \underline{\underline{45.27}} & 46.32 & \underline{\underline{46.05}} \\
\bottomrule
\end{tabular}%
}
\end{table*}

Our extensive evaluation across different environments demonstrates the effectiveness of AMAR in multi-user HAR. Table~\ref{tab:results} presents the precision, recall, and $F_1$-score comparison between AMAR and the benchmarks across three distinct environments: classroom, meeting-room, and empty-room. The results show consistent superior performance of AMAR across almost all settings.

To quantify the overall effectiveness of AMAR, we compute average metrics across all three environments. Our proposed method achieves $F_1$-scores of 53.43\% with RVQ and 56.03\% without RVQ, compared to 45.56\% for the best-performing baseline (DEM-THAT). This represents an improvement of 7.87\% with quantization and 10.47\% without quantization—a relative gain of 17.3\% and 23.0\%, respectively. The performance advantage is even more pronounced for PPS, where AMAR achieves 33.83\% with RVQ and 35.27\% without RVQ, compared to 19.66\% for the best baseline (BCE-THAT). This 1.72$\times$ improvement (nearly doubling the exact set accuracy) demonstrates the substantial benefit of our set prediction formulation over existing approaches.

 The performance gain of AMAR stems from formulating multi-user HAR as a set prediction problem. This is different from BCE-based methods and MultiSenseX that couple the problem with auxiliary tasks (user identity or location) and make independent predictions for each active user. While the DEM approach avoids coupling with auxiliary tasks and achieves competitive precision, it exhibits lower recall, indicating that it overlooks many ongoing activities. 

 We observe that AMAR without RVQ consistently achieves the best performance, while AMAR with RVQ maintains highly competitive results, with the second-best metrics almost always belonging to our quantized approach. The minimal performance gap between the two variants (2.6\% in $F_1$-score, 1.44\% in PPS) demonstrates that RVQ preserves activity-discriminative information despite achieving 99.2\% bandwidth reduction. This validates our residual quantization design, which progressively refines feature representations across multiple layers to minimize information loss while enabling efficient edge-cloud communication.

\subsubsection{Occupancy Counting}
By solving the HAR problem through set prediction, our model automatically estimates the count of people inside the room. This can be thought of as a sub-task that the model must inherently solve to achieve good performance in HAR. Figure~\ref{fig:MAE_count_comprehensive} shows the OCE measured using Eq.~\eqref{eq:MAE_count}, demonstrating the gains of AMAR over other approaches. AMAR achieves substantially lower mean absolute error compared to existing methods, with an average error reduction of 74\% compared to MultiSenseX (which is the second best approach). Specifically, AMAR achieves average OCE of 0.10 across all environments, while MultiSenseX achieves 0.39, and other baselines exhibit errors exceeding much higher. This accurate occupancy estimation validates our set prediction formulation, where the model learns to predict the "no person" class for empty query slots, effectively performing implicit occupancy counting without requiring it as a separate task. We note that while RVQ quantization slightly impacts occupancy counting accuracy (AMAR w/o RVQ achieves 0.05 average OCE), the quantized model still substantially outperforms all baseline methods.
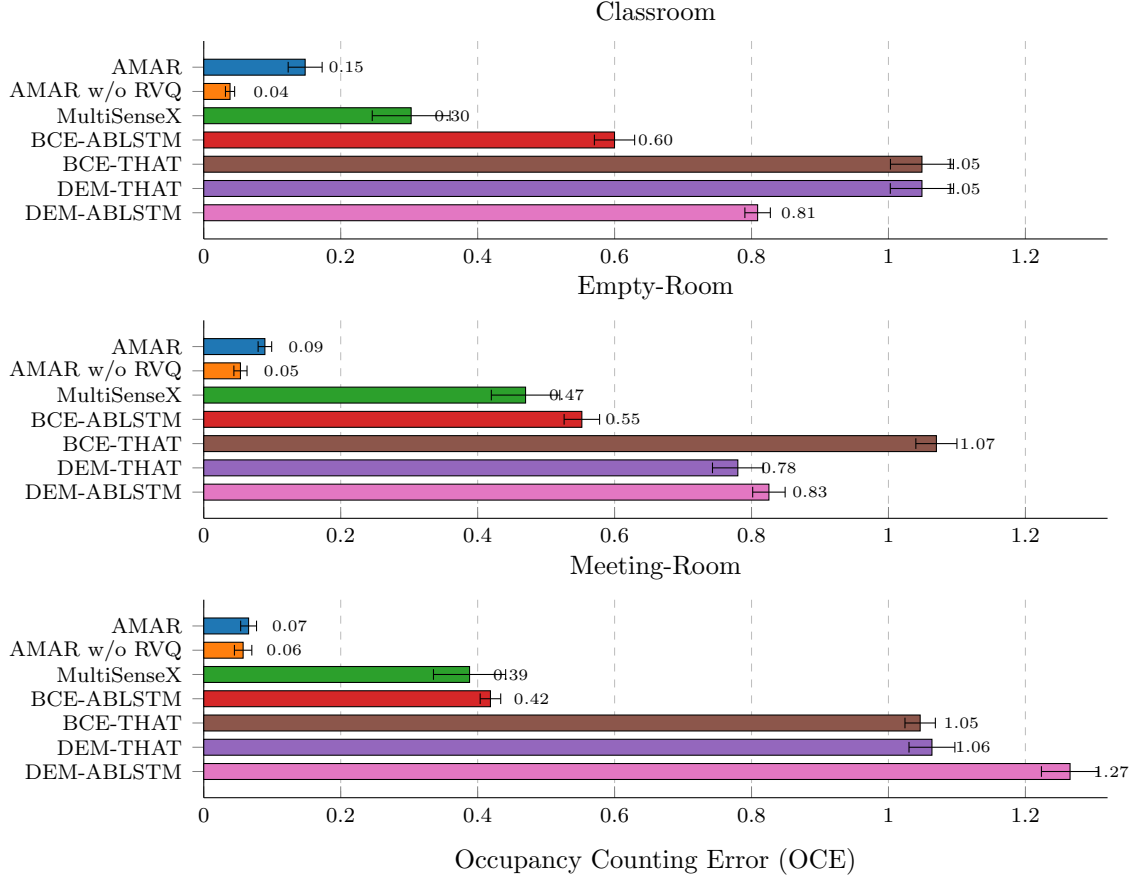
\begin{figure}[t]
\centering
\begin{tikzpicture}

\definecolor{modelcolor1}{HTML}{1f77b4} 
\definecolor{modelcolor2}{HTML}{ff7f0e} 
\definecolor{modelcolor3}{HTML}{2ca02c} 
\definecolor{modelcolor4}{HTML}{d62728} 
\definecolor{modelcolor5}{HTML}{9467bd} 
\definecolor{modelcolor6}{HTML}{8c564b} 
\definecolor{modelcolor7}{HTML}{e377c2} 

\pgfplotsset{
    modelplot/.style={
        nodes near coords,    
        bar shift=0pt,        
        bar width=6pt,        
        error bars/x dir=both, 
        error bars/x explicit   
    }
}

\begin{axis}[
    name=classroom,
    width=0.82\columnwidth,
    height=4.2cm,
    xbar,
    symbolic y coords={AMAR,AMAR w/o RVQ,MultiSenseX,BCE-ABLSTM,BCE-THAT,DEM-THAT,DEM-ABLSTM},
    ytick={AMAR,AMAR w/o RVQ,MultiSenseX,BCE-ABLSTM,BCE-THAT,DEM-THAT,DEM-ABLSTM},
    xlabel={},
    xmin=0,
    xmax=1.32,
    yticklabel style={font=\scriptsize},
    xticklabel style={font=\scriptsize},
    nodes near coords style={font=\tiny, xshift=5pt, /pgf/number format/.cd, fixed, fixed zerofill, precision=2},
    enlarge y limits=0.18,
    axis lines*=left,
    xmajorgrids=true,
    grid style=dashed,
    y dir=reverse,
    title={Classroom},
    title style={font=\small, yshift=-2pt},
    xtick={0,0.2,0.4,0.6,0.8,1.0,1.2},
    xlabel style={font=\small},
    ]
\addplot[modelplot, fill=modelcolor1] coordinates {(0.1481,AMAR) +- (0.0248,0)};
\addplot[modelplot, fill=modelcolor2] coordinates {(0.0384,AMAR w/o RVQ) +- (0.0067,0)};
\addplot[modelplot, fill=modelcolor3] coordinates {(0.3029,MultiSenseX) +- (0.0568,0)};
\addplot[modelplot, fill=modelcolor4] coordinates {(0.5999,BCE-ABLSTM) +- (0.0294,0)};
\addplot[modelplot, fill=modelcolor5] coordinates {(1.0489,DEM-THAT) +- (0.0462,0)};
\addplot[modelplot, fill=modelcolor6] coordinates {(1.0489,BCE-THAT) +- (0.0460,0)};
\addplot[modelplot, fill=modelcolor7] coordinates {(0.8089,DEM-ABLSTM) +- (0.0186,0)};
\end{axis}

\begin{axis}[
    name=empty,
    at={(classroom.below south west)},
    anchor=north west,
    yshift=-18pt,
    width=0.82\columnwidth,
    height=4.2cm,
    xbar,
    symbolic y coords={AMAR,AMAR w/o RVQ,MultiSenseX,BCE-ABLSTM,BCE-THAT,DEM-THAT,DEM-ABLSTM},
    ytick={AMAR,AMAR w/o RVQ,MultiSenseX,BCE-ABLSTM,BCE-THAT,DEM-THAT,DEM-ABLSTM},
    xlabel={},
    xmin=0,
    xmax=1.32,
    yticklabel style={font=\scriptsize},
    xticklabel style={font=\scriptsize},
    nodes near coords style={font=\tiny, xshift=5pt, /pgf/number format/.cd, fixed, fixed zerofill, precision=2},
    enlarge y limits=0.18,
    axis lines*=left,
    xmajorgrids=true,
    grid style=dashed,
    y dir=reverse,
    title={Empty-Room},
    title style={font=\small, yshift=-2pt},
    xtick={0,0.2,0.4,0.6,0.8,1.0,1.2},
    xlabel style={font=\small},
    ]
\addplot[modelplot, fill=modelcolor1] coordinates {(0.0893,AMAR) +- (0.0099,0)};
\addplot[modelplot, fill=modelcolor2] coordinates {(0.0536,AMAR w/o RVQ) +- (0.0096,0)};
\addplot[modelplot, fill=modelcolor3] coordinates {(0.4700,MultiSenseX) +- (0.0500,0)};
\addplot[modelplot, fill=modelcolor4] coordinates {(0.5522,BCE-ABLSTM) +- (0.0259,0)};
\addplot[modelplot, fill=modelcolor5] coordinates {(0.7800,DEM-THAT) +- (0.0370,0)};
\addplot[modelplot, fill=modelcolor6] coordinates {(1.0700,BCE-THAT) +- (0.0300,0)};
\addplot[modelplot, fill=modelcolor7] coordinates {(0.8254,DEM-ABLSTM) +- (0.0237,0)};
\end{axis}

\begin{axis}[
    name=meeting,
    at={(empty.below south west)},
    anchor=north west,
    yshift=-18pt,
    width=0.82\columnwidth,
    height=4.2cm,
    xbar,
    symbolic y coords={AMAR,AMAR w/o RVQ,MultiSenseX,BCE-ABLSTM,BCE-THAT,DEM-THAT,DEM-ABLSTM},
    ytick={AMAR,AMAR w/o RVQ,MultiSenseX,BCE-ABLSTM,BCE-THAT,DEM-THAT,DEM-ABLSTM},
    xlabel={Occupancy Counting Error (OCE)},
    xmin=0,
    xmax=1.32,
    yticklabel style={font=\scriptsize},
    xticklabel style={font=\scriptsize},
    nodes near coords style={font=\tiny, xshift=5pt, /pgf/number format/.cd, fixed, fixed zerofill, precision=2},
    enlarge y limits=0.18,
    axis lines*=left,
    xmajorgrids=true,
    grid style=dashed,
    y dir=reverse,
    title={Meeting-Room},
    title style={font=\small, yshift=-2pt},
    xtick={0,0.2,0.4,0.6,0.8,1.0,1.2},
    xlabel style={font=\small},
    ]
\addplot[modelplot, fill=modelcolor1] coordinates {(0.0655,AMAR) +- (0.0116,0)};
\addplot[modelplot, fill=modelcolor2] coordinates {(0.0575,AMAR w/o RVQ) +- (0.0127,0)};
\addplot[modelplot, fill=modelcolor3] coordinates {(0.3882,MultiSenseX) +- (0.0527,0)};
\addplot[modelplot, fill=modelcolor4] coordinates {(0.4187,BCE-ABLSTM) +- (0.0150,0)};
\addplot[modelplot, fill=modelcolor5] coordinates {(1.0635,DEM-THAT) +- (0.0335,0)};
\addplot[modelplot, fill=modelcolor6] coordinates {(1.0463,BCE-THAT) +- (0.0222,0)};
\addplot[modelplot, fill=modelcolor7] coordinates {(1.2652,DEM-ABLSTM) +- (0.0418,0)};
\end{axis}

\end{tikzpicture}
\caption{OCE comparison across all methods and environments with error bars showing standard error. Lower values indicate better occupancy estimation. Our approach (with and without RVQ)  achieves the lowest errors across all environments.}
\label{fig:MAE_count_comprehensive}
\end{figure}

\subsubsection{Computational Complexity}
Given the edge deployment constraints, we evaluate our model's computational footprint against existing HAR—single user or multi-user—approaches using two key metrics: Parameters (memory requirements) and FLOPs (inference cost).  Table~\ref{tab:computational_complexity} shows that our complete model requires only 0.32M parameters and 501.6M FLOPs, achieving substantial reductions compared to baselines—13.4$\times$ fewer parameters than ABLSTM and 15.3$\times$ fewer than THAT. Notably, AMAR maintains competitive computational efficiency even compared to specialized lightweight architectures, requiring only 4.3$\times$ fewer parameters than BLTHAT while achieving superior recognition performance. In terms of FLOPs, our model demonstrates 6.4$\times$ reduction compared to ABLSTM and 3.6$\times$ reduction compared to THAT, while remaining comparable to CNN-based approaches.
Our edge-cloud architecture provides additional benefits, with the backbone component requiring only 0.11M parameters (34\% of total model complexity) and 177.62M FLOPs for edge deployment. This lightweight design enables real-time CSI processing on resource-constrained devices while maintaining high accuracy through the quantization and cloud-based transformer processing.

\begin{table}[t]
\caption{Computational complexity  across different models}
\centering
\label{tab:computational_complexity}
\small
\begin{tabular}{lcc}
\toprule
\textbf{Model} & \textbf{Parameters (M)} & \textbf{FLOPs (M)} \\
\midrule
LSTM\cite{ordonez2016deep} & 1.62 & 972.96 \\
CNN\cite{ordonez2016deep} & 1.90 & 515.89 \\
THAT\cite{THAT} & 4.89 & 1798.56 \\
CLSTM\cite{ordonez2016deep} & 5.38 & 1790.49 \\
ABLSTM\cite{elkelany2023wifi} & 4.29 & 3211.41 \\
BLTHAT\cite{multiUser_light_iot} & 1.39 & 393.99 \\
\midrule
\textbf{AMAR (complete)} & \textbf{0.32} & \textbf{501.6} \\
\textbf{AMAR backbone (edge)} & \textbf{0.11} & \textbf{177.62} \\
\bottomrule
\end{tabular}
\end{table}
\subsubsection{Communication Efficiency}

A critical advantage of our edge-cloud architecture is the substantial reduction in communication bandwidth achieved through RVQ. Without quantization, transmitting the backbone features $\mathbf{Z} \in \mathbb{R}^{\replength \times \embedDim}$ from edge devices to cloud servers requires ${\replength} \times {\embedDim} \times 32$ bits per sample (assuming 32-bit floating-point representation). With $\embedDim=64$, this amounts to $2048$ bits per temporal position. With our RVQ implementation using $\numRVQLayers=4$ layers and $\codebookSize=16$ prototypes per layer, edge devices transmit only discrete indices requiring $\numRVQLayers \times \log_2(\codebookSize) = 16$ bits per temporal position. This yields a 99.2\% bandwidth reduction compared to transmitting unquantized features. This dramatic reduction in communication overhead makes real-time multi-user sensing practical over standard network connections, while our results in Table~\ref{tab:results} demonstrate that recognition performance remains competitive with the unquantized approach. The minimal performance gap between AMAR with and without RVQ validates that multi-layer RVQ preserves activity-discriminative information.


\subsection{Ablation Studies}

\subsubsection{Effect of the Number of Queries}
As discussed in Section~\ref{sec:opt_objective}, our model requires a fixed number of queries $\numqueries$ that represents the maximum supported occupancy. A natural question arises: how sensitive is the model's performance to the choice of $\numqueries$, particularly when it significantly exceeds the actual occupancy? 
To investigate this, we conduct experiments with varying $\numqueries$ on the WiMANS dataset, which contains up to 5 concurrent users per sample. Specifically, we evaluate the model with $\numqueries \in \{6, 8, 10, 15\}$. Figure~\ref{fig:num_queries_effect} demonstrates that the model maintains consistent performance across all tested configurations, with PPS remaining stable between 33.2\% and 34.7\%. 
This robustness to $\numqueries$ selection indicates that the model effectively learns to predict "no person" for excess queries without degrading its ability to recognize actual activities. Thus, practitioners can set $\numqueries$  to accommodate anticipated maximum occupancy without sacrificing accuracy.

\begin{figure}[t]
\centering
\begin{tikzpicture}
\begin{axis}[
    width=0.9\columnwidth,
    height=5.5cm,
    xlabel={Number of Queries ($\numqueries$)},
    ylabel={PPS (\%)},
    xmin=5, xmax=16,
    ymin=31, ymax=37,
    xtick={6,8,10,15},
    grid=major,
    grid style={dashed,gray!30},
]

\addplot[
    mark=o, 
    thick, 
    color=blue!80, 
    line width=1.2pt,
    error bars/.cd,
    y dir=both,
    y explicit,
] coordinates {
    (6, 33.597) +- (0, 0.6244)
    (8, 33.2011) +- (0, 0.6538)
    (10, 34.66) +- (0, 0.96)
    (15, 34.19) +- (0, 0.84)
};

\end{axis}
\end{tikzpicture}
\caption{Effect of the number of queries on PPS. Error bars represent standard error. The model performance remains stable across different $\numqueries$ values, demonstrating robustness to  the chosen number of queries.}
\label{fig:num_queries_effect}
\end{figure}
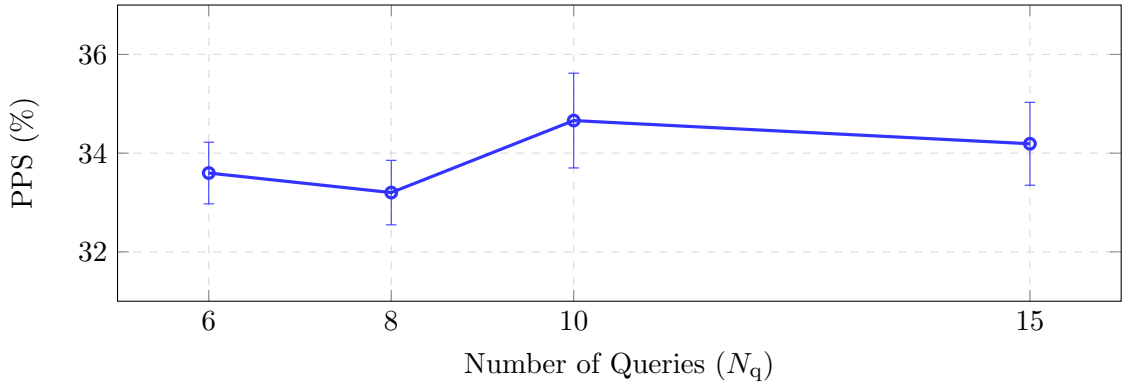

\subsubsection{Effect of RVQ Layers}

We evaluate how RVQ layers affect our algorithm's performance, as quantization typically introduces a trade-off between communication efficiency and recognition accuracy. Figure~\ref{fig:rvq_layers_effect} illustrates the impact of varying both the number of RVQ layers {$\numRVQLayers$} and codebook size {$\codebookSize$} on PPS, with the dashed line representing the model without quantization. The results reveal a consistent pattern across different codebook sizes. Single-layer quantization ($\numRVQLayers=1$) substantially degrades performance compared to the unquantized baseline, but adding more RVQ layers progressively recovers the lost information. As {$\numRVQLayers$} increases from 1 to 6, all configurations approach the no-RVQ baseline, with most achieving near-parity by $\numRVQLayers=6$.

Remarkably, the figure shows that codebook size has relatively modest impact when multiple RVQ layers are employed, with even small codebooks ({$\codebookSize=8$}) achieving competitive performance at higher layer counts. This robustness stems from the exponential growth in representational capacity: with {$\codebookSize=8$} and {$\numRVQLayers=4$}, the system can represent $8^4 = 4,096$ distinct quantization combinations per feature vector, providing sufficient expressiveness to capture complex CSI patterns. Beyond {$\numRVQLayers=6$}, performance stabilizes across all codebook sizes, suggesting that additional layers offer diminishing returns and indicating an optimal balance between representational capacity, model complexity, and communication overhead.
\begin{figure}[t]
\centering
\begin{tikzpicture}
\begin{axis}[
    width=\columnwidth,
    height=6.5cm,
    xlabel={Number of RVQ Layers ($V$)},
    ylabel={PPS (\%)},
    xmin=0, xmax=9,
    ymin=24, ymax=38,
    xtick={1,2,4,6,8},
    grid=major,
    grid style={dashed,gray!30},
    legend style={
        at={(0.97,0.03)},
        anchor=south east,
        font=\small,
        cells={anchor=west}
    },
    legend cell align={left},
]

\addplot[name path=upper, draw=none, forget plot] coordinates {
    (0, 37.26) (9, 37.26)
};
\addplot[name path=lower, draw=none, forget plot] coordinates {
    (0, 34.70) (9, 34.70)
};
\addplot[gray!20, forget plot] fill between[of=upper and lower];

\addplot[thick, dashed, color=black, line width=1.5pt] coordinates {
    (0, 35.98) (9, 35.98)
};
\addlegendentry{w/o RVQ}

\addplot[mark=o, thick, color=blue!80, line width=1.2pt] coordinates {
    (1, 27.05) (2, 31.88) (4, 33.93) (6, 34.85) (8, 35.65)
};
\addlegendentry{$\codebookSize=8$}

\addplot[mark=square, thick, color=red!80, line width=1.2pt] coordinates {
    (1, 27.98) (2, 29.56) (4, 33.79) (6, 35.98) (8, 34.79)
};
\addlegendentry{$\codebookSize=16$}


\addplot[mark=diamond, thick, color=orange!80, line width=1.2pt] coordinates {
    (1, 29.83) (2, 32.80) (4, 33.53) (6, 35.98) (8, 35.25)
};
\addlegendentry{$\codebookSize=64$}

\end{axis}
\end{tikzpicture}
\caption{Effect of RVQ layers on PPS across different codebook sizes. The dashed line shows performance without RVQ (shaded region indicates SE). All RVQ configurations approach the no-quantization baseline as layers increase, demonstrating effective residual quantization with minimal performance loss.}
\label{fig:rvq_layers_effect}
\end{figure}
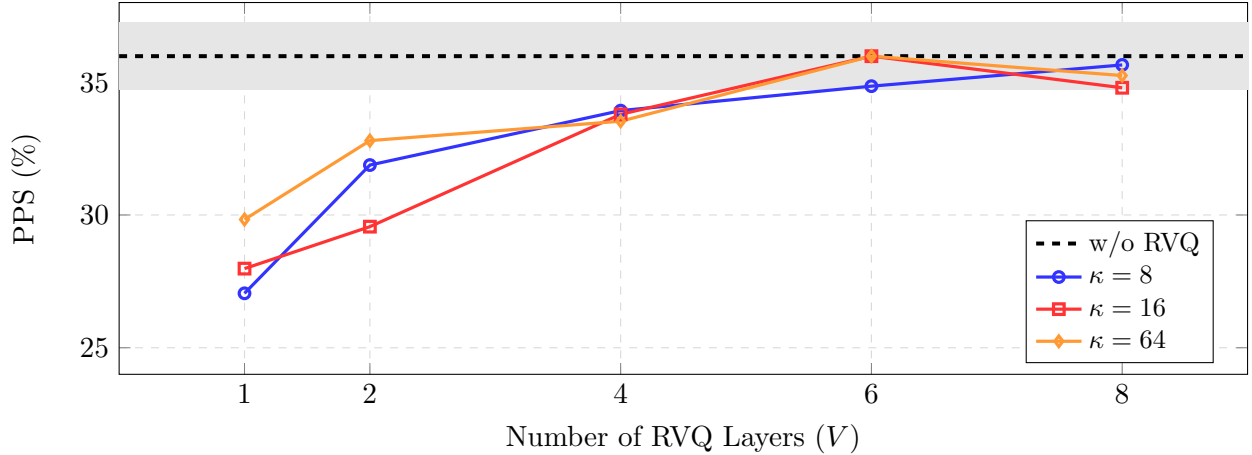

\section{Conclusion}
\label{sec:conclusion}
In this paper, we developed AMAR, a novel framework for inferring multiple concurrent user activities from a single CSI measurement. The framework formulates multi-user HAR as a set prediction problem, which inherently addresses the variable occupancy challenge without requiring prior knowledge of the number of users or specialized hardware configurations beyond standard Wi-Fi infrastructure. This formulation naturally handles the permutation-invariant nature of concurrent activities. To solve the formulated set prediction problem, we designed an end-to-end learnable architecture that employs a transformer-based decoder with learnable query embeddings. These embeddings function as specialized activity detectors, capturing the interdependent nature of overlapping CSI patterns through coordinated attention mechanisms. Recognizing that practical deployment issues, we develop a lightweight backbone network comprising only 0.11M parameters for edge inference, coupled with  RVQ scheme that achieves 99.2\% bandwidth reduction. This edge-cloud split design enables the system to process CSI measurements on resource-constrained devices while maintaining high accuracy through cloud-based transformer processing. Extensive experiments across diverse indoor environments demonstrated that our approach achieves substantial improvements over state-of-the-art methods. 


\bibliographystyle{IEEEtran}
\bibliography{references}

\end{document}